\input{aipcheck}

\documentclass[
    ,final            
  ]
  {aipproc}

\layoutstyle{6x9}

\begin{document}

\title{Precision Holographic Baryons\footnote{An invited talk at Baryons 2010, Osaka, Japan, December 2010}}


\author{Piljin Yi}{
  address={School of Physics, Korea Institute for Advanced Study, Seoul 130-722, Korea}
}

\begin{abstract}
We overview a holographic QCD based on the D4-D8 string theory model,
with emphasis on baryons and nucleon-meson
interactions thereof. Baryons are realized as holographic images
of Skyrmions, but with much qualitative changes. This allows us
to derive, without adjustable parameters, couplings of baryons
to the entire tower of spin one mesons and also to pseudoscalar mesons.
We find some surprisingly good match against empirical values
for nucleons, in particular.
Tensor couplings to all axial-vectors and iso-singlet vectors all vanish,
while, for $\rho$ mesons,  tensor couplings are found to
be dominant. We close with various cautionary comments and speculations.
\end{abstract}

\maketitle


\section{Holographic QCD}

Quantum-Chromodynamics (QCD) in its strongly coupled regime has
defied analytical approach for decades. For instance, the large $N_c$
approximation of `t Hooft offered many new insights
\cite{'tHooft:1973jz}, yet does not let us access the central
features of QCD, namely the confinement and the chiral
symmetry breaking. In more phenomenological approaches, such as
the chiral perturbation theory, we
bypass these fundamental questions,  and work on phenomenology of low energy
hadrons directly. However, this is at the heavy price of
introducing many unknown parameters,
several for each variety of physical particles, to be fixed by data. At the most
practical level, the holography allows us to bridge
the gap between QCD and such phenomenological descriptions,
by ``deriving" an approximate theory of color-singlet objects only,
from a large $N_c$ QCD.

Holography originates from a hypothetical
duality between a strongly coupled open string theory and weakly
coupled closed string theories
\cite{Maldacena:1997re,Gubser:1998bc,Witten:1998qj}.
In practice, one can really compute
 the closed string side when it reduces to a
tree-level theory involving a handful of massless or nearly
massless fields. This invariably includes gravity, since all critical
and closed string theory includes massless graviton, and the truncation
can be justified only when the open string sides, or Yang-Mills side,
is strongly coupled. This latter condition amounts to
$\lambda=g_{YM}^2N_c\gg 1$ in terms of the lowest-lying Yang-Mills
sector of the latter \cite{Maldacena:1997re}. With both $N_c$
and $\lambda$ large, we replace the strongly interacting quantum
theory by a weakly interacting gravity coupled to other light
fields that live in some higher dimensional geometry of
particular shape. These light fields are supposed to
represent gauge-singlet physical particles of the original theory,
which are,  for QCD, what we call hadrons.

The most obvious question, as far as real QCD goes, is whether we can trust anything
that is computed in the large $N_c$ and the large $\lambda$ limit. Real QCD would
have $N_c=3$ while  $\lambda$ is often an extra tunable parameter,
which, for our particular model, comes out to be about 17. Whether or not
these are big enough to justify the truncation is, at the moment, answerable
only by comparing against real world data, so that would be final goal for
this talk.

What we can expect and have seen from holographic QCD's are
relationships between quantities that are hitherto unexplainable,
perhaps except those few quantities already computed by lattice QCD. The earliest such examples
involved mass distribution of glueballs \cite{Csaki:1998qr}. Given a single
input data on the lightest glueball, one often finds
uncanny match with lattice simulations.
The next iteration gave us a series of surprising
structures and predictions in the meson sectors, most notably from
D4-D8 models of Sakai and Sugimoto \cite{sakai-sugimoto}. This model not only reproduced
the Chiral Lagrangian with couplings to vectors, with only one tunable
dimensionless parameter, but it also gave a surprising good prediction
on $\eta'$ mass and a beautiful realization of the vector dominance.

In a similar vein, the holographic baryon in the D4-D8 model  \cite{Hong:2007kx,Hata:2007mb}
is such that we can now discuss, in very precise
terms, what are the meson-baryon couplings, how these translate to
nucleon-nucleon potential models, and ultimately how such holographic
models test against real world data \cite{Hong:2007ay,Hong:2007dq,Deuteron}.
It is the plan of this talk to overview these development with some emphasis on vector meson-nucleon
couplings, which show the true strength of this approach by far.

\section{D4-D8 Model and Holographic Baryons}

Given the short time allowed for this talk and considering that the
most of the audience are nuclear physicists with no or very little
exposure to string theory, it is perhaps best to start with a simple
and reasonably self-contained prescription of the model, instead of
a full-blown derivation.

\begin{figure}
  \includegraphics[height=.2\textheight]{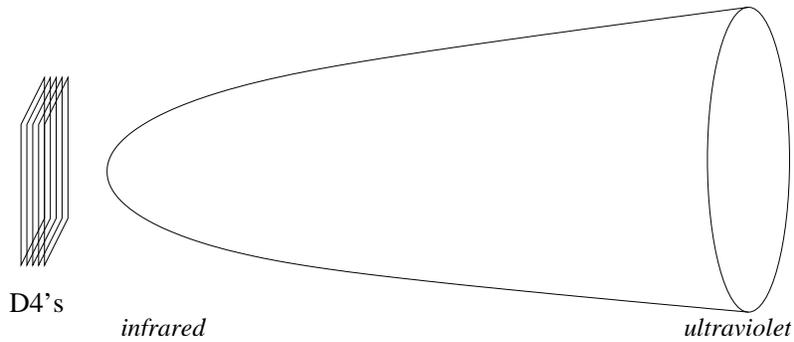}
  \caption{The geometry dual to D4 branes is drawn schematically with both Minkowski
  spacetime and the internal $S^4$ suppressed. $D4$'s are not actually part of the
  spacetime manifold. We only exhibit a cigar-like two-dimensional part of the whole
  ten-dimensional geometry; it encodes how five dimensional supersymmetric Yang-Mills
  theory on D4's is reduced to a four-dimensional nonsupersymmetric QCD.
  The geometry is warped, meaning that excitations localized
  at the bottom (the left end in the figure, denoted as {\it infrared}) are hierarchically
  light, so the lightest of physical particles of QCD live there. The infrared curvature
  scale $\sim M_{KK}$ must be tuned to match a mass scale of QCD.
   }
\end{figure}

For four-dimensional QCD with $N_f$ massless flavors, D4-D8 model
tells us to solve the following five-dimensional $U(N_f)$
gauge theory at tree-level \cite{sakai-sugimoto},
\begin{eqnarray}\label{five}
-\frac14\;\int dx^4dw\;
\; \frac{1}{e(w)^2}\;{\rm tr} {\cal F}^2
+\frac{N_c}{24\pi^2}\int_{4+1}\omega_{5}({\cal A})\: ,\label{dbi}
\end{eqnarray}
where the (weak) position-dependent coupling of this {\it flavor}
gauge theory is
\begin{equation}
\frac{1}{e(w)^2}
=\frac{\lambda N_c}{108\pi^3} \,u(w)\,M_{KK}\: .
\end{equation}
and $\omega_5({\cal A})$ is the Chern-Simons 5-form of $U(N_f)$.
Almost all the dynamical content of this theory is distilled in
the function $u(w)$ that obeys
\begin{equation}
\frac23 \,|w| \,M_{KK}=\int_{1}^u{dy}/\sqrt{y^3-1}\ .
\end{equation}
Note that $u$ ranges from 1 to $\infty$ while $w$ resides in a finite
interval of length $\sim O(1/M_{KK})$. It is clear that $M_{KK}$ sets
all the scales in the meson sector while $\lambda$ serves as additional
dimensionless parameter; it is known that
$M_{KK}\sim 0.94$ GeV is needed to match  the lowest lying $\rho$ meson mass \cite{sakai-sugimoto}.

The origin of this model is in a supersymmetry-broken D4-D8 system in IIA string theory.
The large $N_c$ QCD lives in $N_c$ coincident D4-branes, which, like in any
holographic model, is replaced by its holographic dual geometry \cite{Witten:1998zw}.
The `t Hooft-like coupling can be traced to the underlying string theory as
\begin{equation}
\lambda=2\pi g_s N_c M_{KK} l_s \;,
\end{equation}
where $g_s$ is the string coupling and $1/2\pi l_s^2$ is the tension
of the fundamental string. By analyzing the gravitational physics in
this background, researchers claimed to have found ratios of
glueball masses which are consistent with lattice data within 20\% or so accuracy
\cite{Csaki:1998qr}.

In such a background, $N_f$ D8-branes introduces flavored states \cite{sakai-sugimoto}.
Before taking large $N_c$
limits, open strings connecting D4's and D8's act like quarks, but once D4's are
replaced by its holographic dual geometry, the open strings can only connect D8's and D8's,
and therefore look like bi-quark mesons. The lowest lying sector of these D8-D8 open
string happens to be the five-dimensional $U(N_f)$ gauge theory, from which
one can derive a low energy effective action of pions and spin 1 mesons of infinite
variety. See section 4 for how four-dimensional mesons arise from this flavor gauge theory.

\begin{figure}
  \includegraphics[height=.2\textheight]{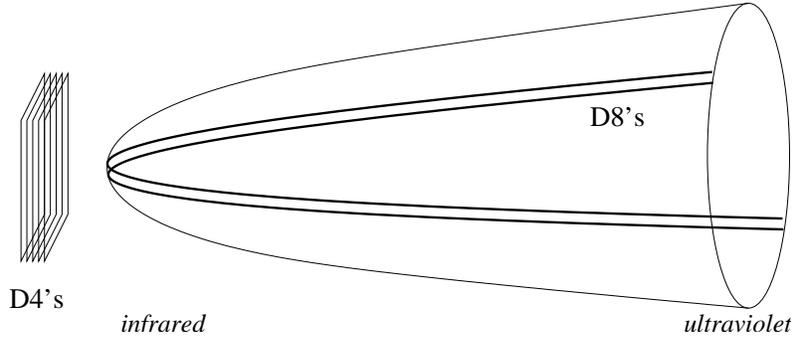}
  \caption{Additional D8-branes carry  a flavor gauge theory in five dimensions,
  consisting of $w$ direction, shown here as the radial direction, and the usual
  four-dimensions of QCD, now shown in this diagram.
  Holographic baryons are coherent state made from this flavor gauge field, of
  unit Pontryagin number, and can also be thought of as a holographic image of Skyrmions.}
\end{figure}

The fact that we found, in the dual holographic description, a flavor gauge
theory is not an accident of this model.
The holography, as can be seen from Witten's original prescriptions \cite{Witten:1998qj},
elevates a continuous global symmetry of the large $N$ field theory in question to a local gauge
symmetry in the holographic description. In fact, the universal presence of
gravity in the holography can be also understood as such a phenomena, since
all relativistic theory possesses the global Poincare symmetry, whose local
form is the diffeomorphism invariance. In the present case, $U(N_f)$ can be
considered as the local form of $U(N_f)_V$ which is the diagonal subgroup
of the usual $U(N_f)_L\times U(N_f)_R$ chiral symmetry. The other, axial
$U(N_f)_A$ does not elevate to a gauge symmetry, precisely because QCD
breaks it dynamically.

This observation is important for us since it also tells us how to find baryons. The
baryon carries $N_c$ unit of $U(1)_V\subset U(N_f)_V$ global quantum number,
also known as the quark number. Denoting $SU(N_f)$ part of the flavor gauge
field by $A$ and its field strength $F$, one can see that the second term
in the five-dimensional flavor theory action includes
\begin{equation}
\frac{N_c}{24\pi^2}\int_{4+1}\omega_{5}({\cal A})\simeq
\frac{N_c}{8\pi^2}\int_{4+1}{\cal A}^{U(1)}\wedge{\rm tr} { F}^2+\cdots \ ,
\end{equation}
so a soliton that carries a unit Pontryagin number,
$\int_{4}{\rm tr} {F}^2=8\pi^2$, carries a unit baryon number also.
In other words, the baryon is a coherent state of ${\cal A}$
with a unit topological winding number in $\pi_3(U(N_f))=\pi_3(SU(N_f))$.

Although similar in spirit to old Skyrmion picture of the baryon \cite{skyrme},
there are qualitative differences. In view of what we will see later when
we reduce the theory to four dimensions, the simplest way to describe the connection
is to say that the holographic baryon is the Skyrmion corrected heavily
due to couplings to an infinite tower of spin 1 mesons \cite{Hong:2007ay}.
 This
difference is important for the phenomenological reasons also, since this new
picture allows us to compute how the baryon would couple to all spin 1 mesons,
resulting in an infinite number of predictions. Let us proceed and see how this
comes about.

\section{An Effective Theory of Holographic Baryons}

Once we identify the baryon as a solitonic object, we must first
try to find a solution, quantize it, and then extract its couplings to
the flavor gauge field. The approximate form of the soliton was
obtained and found to be of very small size \cite{Hong:2007kx,Hata:2007mb}
\begin{equation}\label{size}
\rho_{baryon}\simeq
\frac{({2\cdot 3^7\cdot\pi^2/5})^{1/4}}{M_{KK}\sqrt\lambda }\ ,
\end{equation}
when $\lambda\gg 1$. (If we had truncated to the pion sector only,
upon dimensional reduction, we would have found a Skyrmion of size
$\sim 1/M_{KK}$ instead.)
 It is important to mention here that this
size is not what one typically measures in experiment by lepton
scattering, since the latter is dominated by the so-called vector
dominance and remains finite even as $\lambda\rightarrow \infty$.

Small soliton size is usually a bad sign, since quantum
fluctuation of order Compton length might invalidate the
classical picture altogether. However, the mass of the baryon
scales as $M_{KK}N_c\lambda$, so the Compton wavelength is
comfortably smaller than the soliton size. On the other hand,
the soliton size is still considerably smaller than $1/M_{KK}$,
which as we will see later is the length scale of the
holographic mesons in this model.

Combined, these two facts allows a very simple method for
finding an effective action for the holographic baryon.
The smallness relative to meson mass scale, $\sim M_{KK}$,
implies that as long as the meson interaction goes the
baryon can be treated as if it is a point-like object.
On the other hand, the Compton length's even smaller size
implies that we can take the classical shape of the soliton
seriously.

For simplicity, we take $N_f=2$, in which
case, assuming odd $N_c$, the quantization of the soliton gives
a Dirac particle in the fundamental representation of $SU(N_f=2)$
as the lowest lying quantum state. Generalization to higher isospin baryons
is straightforward if somewhat more involved. See Ref.~\cite{Park:2008sp} for
the formulation and Ref.~\cite{Grigoryan:2009pp} for an application.
Introducing a five-dimensional
Dirac field ${\cal B}$ of isospin 1/2, we arrive at
an effective nucleon action with couplings to the flavor gauge field \cite{Hong:2007kx,Hong:2007ay},
\begin{eqnarray}
&&\int d^4 x \,dw\left[-i\bar{\cal B}\gamma^m D_m {\cal B}
-i m_{\cal B}(w)\bar{\cal B}{\cal B} +{2\pi^2\rho_{baryon}^2\over
3e^2(w)}\bar{\cal B}\gamma^{mn}F_{mn}{\cal B} \right] \ .
\label{5d}
\end{eqnarray}
The position-dependent soliton mass is $m_{\cal B}(w)= 4\pi^2/e(w)^2$
in the leading $1/N_c$ approximation.
The interaction with mesons
are captured in two  couplings to the gauge field.
The first is via the covariant derivative
\begin{equation}
D_m\equiv \partial_m-i(N_c {\cal A}_m^{U(1)}+{A}_m) \ ,
\end{equation}
for which the flavor gauge field  ${\cal A}_m$ of (\ref{five})
is decomposed  as ${\cal A}_m^{U(1)}+A_m$ with
traceless $2\times 2$ $A_m$.
This is entirely determined by the conserved quantum
numbers of the baryon.

The second, less familiar looking
term couples the baryon directly to the $SU(2)$ field strength $F$.
One might mistake that we find this term as a first order correction
term, via usual effective
theory argument based on operator dimension counting. This is
not so, however. This ostensibly higher order term is
actually the leading $1/N_c$ coupling to $SU(N_f=2)$
part of the flavor gauge field. Although we do not elaborate on
the derivation here, we emphasize that not only the structure of
the operator $\bar{\cal B}\gamma\cdot F{\cal B}$ (which happens to be
also unique at dimension six level) but also the coefficient
function have been derived rigorously \cite{Hong:2007kx}.
The derivation relies on the instanton-like nature of the
soliton, and shares a mathematical reasoning with
the well-known treatment of the Skyrmion by Adkins, Nappi,
and Witten \cite{ANW}.

The coefficient function  of $\bar {\cal B}\gamma\cdot F{\cal B}$
is actually computable unambiguously only for $w=0$.
Possible deviation would be a multiplicative correction that is unit at $w=0$.
Thankfully, this does not affect large $N_c$ estimates of couplings we
consider, because the baryon wavefunction along $w$ is very tightly
localized at $w=0$ due to the strongly confining nature of $m_{\cal B}(w)$.

\section{4D Physics}

Actions (\ref{five}) and (\ref{5d}) are meant to be used to generate
tree-level Feynman diagrams for mesons and (isospin 1/2) baryons, but they are still
in the five-dimensional form. What we need to do at this stage is to
reduce the system by performing a Kaluza-Klein reduction along the holographic $w$-direction.
This is exactly what was done to extract glueball masses \cite{Csaki:1998qr}.

For the flavor  gauge field, we mode-expand \cite{sakai-sugimoto,Sakai:2005yt}
with eigenfunctions, $\psi_{(n)}(w)$,
\begin{equation}\label{ex}
{\cal A}_\mu(x;w)= i\,[U^{-1/2},\partial_\mu U^{1/2}]/2 +i\{U^{-1/2},\partial_\mu U^{1/2}\}\psi_0(w)
+\sum_n {\cal A}_\mu^{(n)}(x)\psi_{(n)}(w)\:
\end{equation}
with $
U(x)=exp({i\int_w {\cal A}})$. The gauge kinetic term in (\ref{five})
produces two type of terms. The first, from $\psi_{(n)}$'s, contains
an infinite tower of vector and axial-vector mesons,
\begin{equation}
\int dx^4 \sum_{n=1}^\infty{\rm tr}\; \left\{{1\over 2} {\cal F}_{\mu\nu}^{(n)}
{\cal F}^{(n)\mu\nu}+m_{(n)}^2 {\cal A}_\mu^{(n)} {\cal A}^{(n)\mu}\right\}+\cdots\:,
\end{equation}
with the eigenvalue $m_{(n)}^2$ of the KK mode $\psi_{(n)}$, and
${\cal  F}^{(n)}_{\mu\nu}=\partial_\mu {\cal A }^{(n)}_\nu-\partial_\nu
{\cal A}^{(n)}_\mu$. Because the only length scale in this problem of KK-reduction is
$1/M_{KK}$ the masses have the form $m_{(n)}^2\sim b_nM_{KK}^2$ with a
discrete tower of positive numbers $b_n$'s that start at $0.6$ or so.
We suppressed interaction terms here but wish to emphasize that all the
couplings as well as the masses are unambiguously fixed, once we fix
$\lambda$ and $M_{KK}$. The meson sector requires
$M_{KK}\simeq 0.94$ GeV, to fit the lightest $\rho$-meson mass, and
$\lambda\simeq 17$, to fit $f_\pi$ to the pion decay constant.

Making contact with four-dimensional spin 1 mesons is done by
\begin{equation}
{\cal A}_\mu^{(2k-1)}=
\omega^{(k)}_\mu\left(\begin{array}{cc}1/2 &0 \\ 0&1/2\end{array}\right)+\rho_\mu^{(k)a}\frac{\tau^a}{2}\:,
\end{equation}
for vectors and by
\begin{equation}
{\cal A}_\mu^{(2k)}=
f^{(k)}_\mu\left(\begin{array}{cc}1/2 &0 \\ 0&1/2\end{array}\right)+a_\mu^{(k)a}\frac{\tau^a}{2}\:,
\end{equation}
for axial-vectors. Thus, iso-triplets are denoted as $\rho$'s and $a$'s while singlets
are denoted as $\omega$'s and $f$'s. Even/odd nature of $\psi_{(n)}(w)$ translates to the
usual parity of the corresponding mesons, so $\rho$'s and $\omega$'s are vectors
and $a$'s and $f$'s are axial vectors, appearing alternately mass-wise.

The Chiral Lagrangian of Goldstone bosons from the broken
chiral symmetry is the other part, from the zero mode $\psi_0$,
\begin{equation}\label{Skyrme}
+\int dx^4 \;\left({f_\pi^2\over 4}{\rm tr} \left(U^{-1}\partial_\mu
U\right)^2 +{1\over 32 e^2_{Skyr
me}} {\rm tr} \left[ U^{-1}\partial_\mu U,
U^{-1} \partial_\nu U \right]^2\right)\:,
\end{equation}
with $ f_\pi^2=\lambda N_c M_{KK}^2/54\pi^4$ and
$1/e^2_{Skyrme}\simeq  {61 \lambda N_c}/54\pi^7$. $U$ is related to
pions and $\eta'$ (meaning the $U(1)$ part, regardless of $N_f$) as
\begin{equation}
U=exp(\pi i (\eta'+\pi^a\tau_a)/f_\pi)
\end{equation}
The Chern-Simons terms generate additional interaction term,
and among them is the Wess-Zumino-Witten term.

For baryons, let us keep only the nucleons for
simplicity,
corresponding to the lowest-lying
Kaluza-Klein mode, by declaring ${\cal B}_\pm(x^\mu,w)={\cal
N}_\pm (x^\mu)f_\pm (w)$, with $\gamma_5{\cal N}_\pm=\pm{\cal N}_\pm$.
 Solving the mode equations for the smallest eigenvalue
$m_{\cal N}\simeq m_{\cal B}(0)+O(M_{KK})$,
\begin{eqnarray}
&&\left[-\partial^2_w \mp \partial_w m_{\cal B}(w)+(m_{\cal B}(w))^2\right]
f_\pm(w)=m_{\cal N}^2 f_\pm(w)\:,\label{eigeneq}
\end{eqnarray}
and reconstituting ${\cal N}_\pm$ into a single
four-dimensional nucleon field ${\cal N}$,  we find from (\ref{5d}) the following
 four-dimensional nucleon action \cite{Hong:2007kx,Hong:2007ay,Deuteron},
\begin{equation}
\int dx^4\;{\cal L}_4 = \int dx^4\left(-i\bar {\cal N}
\gamma^\mu\partial_\mu {\cal N}-im_{\cal N}\bar {\cal N}{\cal N}+ \cdots\right)\:,
\end{equation}
where the ellipsis denotes all couplings between nucleons and mesons. These
include cubic couplings to chiral Goldstone bosons and  axial-vector mesons
\begin{eqnarray}
\frac{1}{2f_\pi}\bar {\cal N}  \gamma^\mu\gamma^5
\left[g_A\partial_\mu\pi^a\tau_a+g_{\eta'}\partial_\mu\eta' \right]  {\cal N}
-\frac12
\bar {\cal N} \gamma^\mu\gamma^5 \sum_{k\ge 1}\left[g_f^{(k)} f^{(k)}_\mu+g_a^{(k)} a_\mu^{(k)a}{\tau^a}\right] {\cal N}\ ,
\end{eqnarray}
and dimension-four and -five cubic couplings to vector mesons
\begin{eqnarray}
-\frac12 \bar {\cal N} \gamma^\mu
 \sum_{k\ge 1}\left[g_{\omega}^{(k)} \omega_\mu^{(k)}+g_{\rho}^{(k)}\rho^{(k)a}_\mu\tau_a\right]
 {\cal N}
 +\frac12
\bar {\cal N} \gamma^{\mu\nu}
 \sum_{k\ge 1} \left[ g_{d\rho}^{(k)}\partial_\mu\rho_\nu^{(k)a}\tau_a\right]
 {\cal N}\ .\label{vector-coupling}
\end{eqnarray}
All coupling constants are computed by  overlap integrals of type
$\int dw\;f_+^*f_\pm \psi_{(n)}$, possibly with a derivative $\partial_w$
acting of $\psi_{(n)}$. In the large $\lambda N_c$ limit, the baryon
wavefunctions $f_\pm$ are very sharply peaked at $w=0$, relative to $\psi_{(n)}$'s,
which makes its bilinears act like a delta function. This allows
leading large $N_c$ contributions to these couplings be estimated unambiguously.

The last, dimension-five derivative couplings deserve further comments.
Note that, among spin 1 mesons, only iso-triplet vectors, denoted as $\rho^{(k)}$,
have such derivative-couplings to nucleons. None of axial vectors (iso-singlet $f$ or
iso-triplet $a$), nor any of iso-singlet vectors $\omega$, have a coupling of
this type. Interestingly, for $\rho$ mesons, these tensor couplings are
dominant over the usual vector couplings. As we will see, for $\rho^{(1)}$
and $\omega^{(1)}$, much of what we find here have been known empirically
and used crucially in potential models of nucleons. These are perhaps
the most striking results out of D4-D8 holographic baryon picture.

\section{Numbers and Comments}

As we noted early on, it is unclear to what extent we should
take predictions of such a holographic QCD seriously.
A priori, there seems to be more reason not to do so, especially
when we talk about the baryon sector which is very heavy in the
large $N_c$ limit. QCD, like any asymptotically-free field theory,
comes with a natural scale where the theory become strongly coupled,
which in this model is a fraction of $M_{KK}$. On the other hand,
the fact that the holographic QCD is truncated to gravity and its
immediate low energy partners implies another scale associated with
this truncation, which is in this model nothing but $M_{KK}$.
So, while very low energy sector of such a description may be
trustworthy, we appear to have little justification for extending the
description beyond 1 GeV.
Nevertheless, we find  results which mimic experimental findings
remarkably well. In fact, it has been explained, based on string
theory computations, how the baryon sector of this holographic
QCD model can produce as competitive results as the meson sector.
We refer the audience to the relevant literature \cite{Yi:2009et,Hashimoto:2010je}.

Probably the most striking and robust prediction of this model is
the absence of dimension-five derivative couplings (or tensor couplings)
to iso-singlet $\omega$ mesons as well as to those of
axial-vector mesons \cite{Deuteron}
\begin{equation}
g^{(k)}_{d\omega}=0\ ,\qquad g^{(k)}_{df}=0\ ,\qquad g^{(k)}_{da}=0 \ ,
\end{equation}
as we already noted. This is true as long as
we start with (\ref{5d}) and extract the couplings at tree-level.

Remarkably, $g^{(1)}_{d\omega}=0$ reproduces an empirical fact  \cite{tensor}
that, without any theoretical understanding at all,
has been used by nucleon-nucleon potential models for decades. In this holographic
approach, this vanishing of $g_{d\omega}$'s happens simply because of the
instanton-like nature
of the baryon, or, in terms of (\ref{5d}), because the derivative
couplings exist only for $SU(2)$ part of the flavor gauge field.
For axial mesons,
which are all heavier than nucleons in real world QCD,
empirical values for tensor  couplings appear
unavailable. It would be most interesting if $g^{(k)}_{df}=0=g^{(k)}_{da}$
can also be verified experimentally.

In contrast, $g^{(k)}_{d\rho}$'s  do not vanish but are dominant
over the dimension-four vector coupling in the large $N_c$ limit \cite{Deuteron} as
\begin{equation}
2M_{KK}\times g^{(k)}_{d\rho} \simeq 1.3\times {N_c}\times g^{(k)}_\rho \ .
\end{equation}
After extrapolating to $N_c=3$ and $\lambda\simeq 17$  and adding a subleading
correction, we find
\begin{equation}
2M_{KK}\times g^{(1)}_{d\rho} \simeq 6.0 \times g^{(1)}_\rho \ .
\end{equation}
If\footnote{Unfortunately,  prediction
of the baryon mass scale relative to the meson one remains the most
problematic aspect of this model. The most naive estimate for
$m_{\cal N}$ is too large by a factor of 2, although one must eventually
consider summing up all bosonic and fermionic modes around the soliton.}
$m_{\cal N}\simeq M_{KK}$ as the real world nucleon mass suggests, this
would coincide with the empirical ratio between the tensor
and the vector coupling, found and used in nucleon potential
models \cite{tensor}, which again had no theoretical understanding at all.
In particular, this is, from typical low energy effective theory
viewpoint, all the more surprising since it tells us that higher dimensional
operators can be more important in theories like QCD with its
low intrinsic energy scale.

Another set of robust results concern ratios between iso-singlet
and iso-triplet couplings. There is a universal relationship of
type \cite{Hong:2007ay}
\begin{equation}
{g_{\omega}^{(k)}}/{g_{\rho}^{(k)}}\simeq N_c+\delta(k) \;,
\end{equation}
where the subleading correction $\delta(k)$ turns out to be
positive. Again, extrapolating to real QCD regime, we find for $k=1$
\begin{equation}
{g_{\omega}^{(1)}}/{g_{\rho}^{(1)}}\simeq 3+0.6 =3.6 \;.
\end{equation}
Extracting ratios like this empirically from experiments is
fairly model-dependent, but the ratio is believed to be
larger than 3 and numbers around 4 to 5 are typically found
\cite{hoehler-pietarinen}.
Given the roughness of the approximation involved here,
this result is also remarkably good.

Our prediction for $g_A$, whose experimental value is about 1.27, is \cite{Hong:2007kx}
\begin{equation}\label{gA}
g_A=
\left(\frac{24}{5\pi^2}\right)^{1/2}\times\frac{N_c}{3}+\cdots \;.
\end{equation}
With a subleading correction added, the extrapolated value is 0.84
which is about 30\% better than Adkins-Nappi-Witten estimate \cite{ANW} based
on the vanilla Skyrmion but worse than an improved Skyrmion dressed
by the lowest lying $\rho$ meson with phenomenologically
determined couplings \cite{vector-skyrmion}. It has been
argued that  there is a subleading
group theoretical correction that shift $N_c\rightarrow N_c+2$
in (\ref{gA}), upon which we find  about 1.3. Validity of this
claim remains disputed.

We wish to close with a caveat. Comparisons with experiment must involve direct
computation of scattering amplitudes within the model, since,
more often than not, empirical values of coupling are somewhat
model-dependent. Much work needed to be done here.
Nevertheless, it seems that
D4-D8 holographic QCD has proven to be very a predictive and successful
model of QCD, despite many potential problems in extrapolating to the
realistic regime.  The baryon sector is particularly difficult to justify,
yet produced many robust results that agree with real world observations.
This is particularly a pleasant surprise, given the fact that the model has
only one dimensionless tunable parameter is.

A holographic model of
any known variety cannot hope to compete against various low
energy models of mesons and nuclei, for the latter comes with large
number of tunable parameters. The main reason for studying holographic
QCD model would not be in producing models that are competent in this sense,
but rather in advancing our understanding of QCD proper.
We hope that numbers presented in this note would eventually
prove to be useful guiding lights.

Finally, we must also caution
the audience that the model is expected to fail dramatically at very high
energy, say, several GeV and beyond, since it does not exhibit the
correct Asymptotic Freedom. This is no different from why
we do not rely on the Chiral perturbation theory in to study QCD
in the perturbative regime. The asymptotic freedom remains among a wish list that
must be fulfilled to achieve a true holographic QCD built from
reliable and true string theory solutions.

\begin{theacknowledgments}
I would like to thank organizers of Baryons 2010 for invitation and also
for the stimulating environment during the conference.
P.Y. is supported in part by the National Research Foundation of Korea (NRF)
via Basic Science Research Program (grant number 2010-0013526).
\end{theacknowledgments}



\bibliographystyle{aipproc}   

\end{document}